# Design of a Smart Waste Management System for the City of Johannesburg

Beauty L. Komane [1] and Topside E. Mathonsi [2]

[1,2]Department of Information Technology, Faculty of Information and Communication Technology, Tshwane University of Technology, Pretoria, South Africa
`komanebeauty802@gmail.com` and
`MathonsiTE@tut.ac.za`

**Abstract.** Every human being in this world produces waste. South Africa is a developing country with many townships that have limited waste resources. Over-increasing population growth overpowers the volume of most municipal authorities to provide even the most essential services. Waste in townships is produced via littering, dumping of bins, cutting of trees, dumping of waste near rivers, and overrunning of waste bins. Waste increases diseases, air pollution, and environmental pollution, and lastly increases gas emissions that contribute to the release of greenhouse gases. The ungathered waste is dumped widely in the streets and drains contributing to flooding, breeding of insects, rodent vectors, and spreading of diseases. Therefore, the aim of this paper is to design a smart waste management system for the city of Johannesburg. The city of Johannesburg contains waste municipality workers and has provided some areas with waste resources such as waste bins and trucks for collecting waste. But the problem is that the resources only are not enough to solve the problem of waste in the city. The waste municipality uses traditional ways of collecting waste such as going to each street and picking up waste bins. The traditional way has worked for years but as the population is increasing more waste is produced which causes various problems for the waste municipalities and the public at large. The proposed system consists of sensors, user applications, and a real-time monitoring system. This paper adopts the experimental methodology.

## 1 Introduction

One of the greatest challenges facing developing countries is the unhealthy disposal of solid waste which resulted from human activities of development and survival. The awareness and attitude of people in the community appear to be crucial. The problem of waste management has arisen recently in developing countries where there is a little history of the implementation of formal and informal community environmental education awareness programs. The solid waste problem is a continually increasing factor worldwide [1]-[8]. *Khan et al. [1]* stated that waste management is one of the biggest concerns in our day-to-day livelihoods and natural environment. The traditional method of waste management carries lacking in some areas such as garbage overflows causing environmental pollution and unhygienic living conditions.

Besides the amount of time and manpower required in this field is broad. Solid waste produced by people from society and animals includes – papers, old clothes, food, sanitary pads, used tissues, trees, plastics, organic and inorganic waste. The increasing disposal of solid waste affects the world in many components including – air pollution, environmental pollution, diseases, and an unhygienic lifestyle. *Pratima [8]* stated that waste also causes, water pollution, producing of gas emissions, Groundwater contamination by the leachate generated by the waste dump, Surface water contamination by the run-off from the waste dump, Bad odour, pests, rodents and wind-blown litter in and around the waste dump, Generation of inflammable gas such as methane within the waste dump, Bird menace above the waste dump which affects the flight of aircraft, Fires within the waste dump, Erosion and stability problems relating to slopes of the waste dump, Epidemics through stray animals, Acidity to surrounding soil and Release of greenhouse gas. This paper is aimed to design a smart waste management system in the city of Johannesburg. This paper will help in managing waste smart by integrating factors ranging from the storage, collection, transfer and transport, processing, and disposal stages. The most important factor in the unmannered disposal of waste as proposed by *Pratima* [8] is the risks to the environment and to public health. Studies have shown that a high percentage of workers who handle refuse and of individuals who live near or on disposal sites are infected with gastrointestinal parasites, worms, and related organisms. This paper adopts the experimental methodology. An experimental methodology is chosen because it allows real-time results from the tested parameters.

The adopted system contains a model that is an integration of sensors, this model will be installed inside waste bins at the bottom of the lid, and this model will have sensors that will sense whether the bins are full or not. Secondly, we will have CCTV that will collect real-time information on whether the side dumping sites are full or not, and finally, a monitoring site where all data will be displayed and reported back to the central monitoring centre. Lastly, an application that will be used for the public, this application will let the public know which bin is full, and which is not. So that when they have waste to throw, they can find bins that are not full. This will be advantageous to the municipalities as time and costs will be saved. This solution will be accompanied by education awareness programs for the public on how to use the application and how to be hygienically clean to protect their own health and the environment.

This paper is not to underestimate the work the City of Johannesburg does in dealing with the waste in the city but to extend the knowledge and solutions that can help in managing the waste better including the use of ICT solutions instead of traditional solutions.

The remainder of this paper is organized as follows: In Section II, we provide the related work, where we discuss previously proposed algorithms/solutions used to solve the identified problem. We further provide the gaps when these solutions were applied. We provide an argument and discussion in Section III. In Section IV, we present the proposed solution. We look at education and awareness in Section V. Lastly, we provide a conclusion in Section VI.

## 2 Related Work

*Xenya et al.* [4] stated that waste management in Ghana has been a problem that affects its industrial cities. The supply of waste bins at vantage points by waste management institutions has not been enough to solve the problem of the current management system. They have proposed a technique that contains multiple solutions to address the issue of waste in the area, spillage, and incompetent collection schemes. Their proposed solution is an automated system that could issue regular notifications to appropriate personnel assigned for waste management activities. The proposed system is adopting the concepts of IoT and smart cities by constructing sensor nodes (smart waste bins) that can detect and report the level of solid waste to appropriate nodes for further action. The smart bin top cover is embedded with a micro controller-based circuit with an ultrasonic sensor to detect the level of garbage.

A GSM module is included in the hardware system for internet connectivity to the cloud-based database (Thing Speak) and SMS notifications to the garbage collectors. The system also includes a management system that contains a database of the waste bins with real-time access to the bin status, geolocation of bins, routing information, and a work order form. The system contains a monitoring platform to handle the alert records by creating orders for the garbage collectors/drivers which can be accessed via a mobile application system. They concluded that the system implementation works successfully, making it possible to monitor waste bin status in real time but occasionally had high latency which is primarily due to the use of a GSM module for GSM/GPRS connectivity.

*Jordosh et al*. [6] focus on modernization and advancement in technology that they can use to improve the overall urban waste management system. They argue that the current smart bin separates waste and puts it in the appropriate part, but none has connectivity. They stated that the problem they face as a country the waste disposal, segregation, and collection. Which collectively is one of the biggest challenges faced by the authorities. With the increasing or rather rise in urbanization, there is an urgent need for solid waste management. The initiative undertaken by the current Prime Minister, Shri Narendra Modi, the Swachh Bharat Mission, has made us proactive in developing a solution to this problem. Therefore, they have proposed SEGRO. The main reason as stated by the authors for efficient waste collection, and disposal is for the protection of the environment and the health of the population.

Thus, even if the smart bin is deployed widely, they will not be viable enough, without proper collection facilities for them. They developed a central server that is connected to each of the smart bins and undertakes all the necessary functions such as tracking all the bins for their garbage level, tracking all the trucks, and generating optimized routes for the truck drivers. They added that the recycling industry also is disorganized. Sending the collected waste directly to these industries helps in saving time, fuel, and storage space. They concluded that a consolidated system that serves as an end-to-end product for waste management is the solution they are proposing. SEGRO is an end-to-end solution, encompassing waste disposal, segregation, collection, and the effective usage of recycled waste. They propose a solution that will not only effectively separate the waste into different types based on their life cycle but also will provide an optimal algorithm for the collection of waste.

*Dubey et al. [7]* stated that with the increasing population and industrialization of nations, waste management has become a challenging issue. Small-scale waste management is also adding the same capacity as large-scale waste management. They stated that the IoT and machine learning-based waste management systems for residential society are aimed to enhance the same concern as the waste management of the smart city. Their paper employs monitoring of various dustbins located in different residential societies. The dustbin is equipped with sensors that monitor for dustbin capacity, metal level, and poisonous gas level. Machine learning classification techniques such as SVM, NB, RF, DT, and KNN are used to test their ability to predict the accuracy of sending alert messages to third parties to manage the waste of society. In addition, their results suggest that the RF algorithm produced the most accurate forecasts of the alert message. The accuracy of the RF algorithm is 85.29 %. The overall impact of this research is in the upliftment of green technologies by reducing pollution in the smart city. The hardware used for the proposed framework consists of a Raspberry with having Wi-Fi component. The capacity and metal level of dustbins is monitored using a standard ultrasonic range sensor, and for poisonous gases used MQ4 sensor. They used a Python-based application to retrieve the monitoring data from the sensor and send it to the central monitoring server. The use of Scikit library provides the functionality of machine learning capabilities such as classification algorithms.

Their system consists of a parameter to Monitor Dustbin The system contains a process that involves three parameters for monitoring and analysis. Dustbin level, Metal level, Poisonous gas level are to monitor dustbin capacity level, ultrasonic sensors are used. The ultrasonic sensor below the lid of the dustbin continuously monitors the dustbin for the status of waste inside the dustbin and helps in calculating the distance of waste from the bottom of the dustbin and intimate the central cloud waste management center when the dustbin is 70% filled. The MQ4 sensor fixed inside the dustbin continuously checks or the presence of poisonous gas like methane. This sensor can sense the concentration of poisonous gas in the range 300 ppmto 10000 ppm. Centralized Server: The information from each of the respective bins is collected at specific intervals and stored in the linked server for further processing. DL, ML, and PG are important parameters taken for analysis. The data set is taken in a CSV file.

The machine learning classification technique is used to monitor the real-time bin value of DL, ML, and PG levels and send alert messages to third parties. Their proposed framework has developed a concept to segregate and manage household wastage using IoT and machine learning. In their work, they have taken three parameters for monitoring and analysis of a dustbin. They have used ultrasonic sensors and MQ4 sensors to monitor the dustbin status. In their experiment, they evaluated their proposed method by using KNN, LR, NB, SVM, RF, and DT. The result shows RF accuracy is 85%. The proposed work will also utilize to record the waste statistics and provides the future waste generation rate of residential areas that may be subsequently used to determine the optimum size of the dustbin and recycle the waste.

*Pratima [8]* stated that one of the major challenges facing developing countries is the unhealthy disposal of solid waste which resulted from human activities of development and survival. They outlined that the problem of waste management has risen recently in developing countries where there is little history of implementation of formal and informal community environmental education awareness programs. Information communication technology (ICT) significantly supports the acquisition and amalgamation of knowledge, offering developing countries unprecedented opportunities to enrich educational systems. They supported their theory with the fact that a large amount of waste is disposed of on land in open dumps or in improperly designed landfills. The disposal of waste has many impacts on the environment such as groundwater contamination by the leachate generated by the waste dump, Surface water contamination by the run-off from the waste dump, bad odor, pests, rodents, and wind-blown litter in and around the waste dump and Generation of inflammable gas such as methane within the waste dump. Their study focuses on statistics, a sample is a subject chosen from a population for investigation; a random sample is one chosen by a method involving an unpredictable component. Random sampling can also refer to taking several independent observations from the same probability distribution, without involving any real population.

The sample usually is not representative of the population from which it was drawn. This random variation in the results is termed a sampling error. In the case of random samples, mathematical theory is available to assess the sampling error. Thus, estimates obtained from random samples can be accompanied by measures of the uncertainty associated with the estimate. This can take the form of a standard error, or if the sample is large enough for the central limit theorem to take effect, confidence intervals may be calculated. For data collection, a simple random sampling method is used. A simple random sample is selected so that all samples of the same size have an equal chance of being selected from the population. They concluded that at the end of the research investigation, during the comparison of two data sets of all the participants namely scores before and after applying the method i.e., educating them by means of ICT, when analyzed by t-test showed that the average score of students is increased due to use of computer-aided learning method, signifying the development of effective computer-assisted learning tool as far as the awareness of survey population is concerned.

*Khan et al.* [1] Waste management is one of the biggest concerns in their day-to-day lives and environment. The traditional method that they use is lacking in some areas such as garbage overflows causing environmental pollution and unhygienic living conditions. The amount of time and manpower requires in this field is extensive. This research proposes an advanced method in which waste management is more likely to be automated and saves time as well as cost. Through the use of this advanced method, a Smart Bin has been designed and implemented. Microcontroller is used to form the heart of this system interfacing with sensors to detect waste levels and a GSM for data transmitting and receiving purposes. This proposed system would have an automated waste level detection process and a smart monitoring and overall management process. The technologies that have been used in this system are said to be good enough to prevent garbage overflow and ensure the partial and perfect waste management and monitoring system maintaining a green environment.

The real-time waste level detection and the smart monitoring and recycling system make this method more effective. Their solution consists of a smart bin that contains an electrical device that consists of sensors and a microcontroller and a GSM as the transmitter. Their study uses one kind of sensor which is the ultrasonic sensor to sense the level of waste in the bin. Whenever the bin is full the sensor will detect and will send the information to the zone control unit by GSM. They performed a simulation to reveal whether the circuit will provide the desired output and the errors that needed attention. This paper provides the design process of a smart waste management system and how to build a smart bin that could be used in countries like Bangladesh. Their research focuses on building a simple and cost-effective smart bin. They have concluded with a smart bin with numerous features which can be used in several ways, according to need and thus this design of a smart waste management system and design of smart bin is effective and quite user-friendly.

*Thibuy et al.* [3] stated that recycling waste is valuable waste due to its potential to be used as raw material. According to their observation and data collection at Bangsaen Beach, Saensuk City, Chon Buri Province, Thailand, a lot of valuable waste is being thrown away instead of being collected and separated properly. This leads to not only the raw material problem but also environmental pollution. They found out that the 2 main reasons because people do not litter their waste appropriately and they lack basic knowledge of how to separate the valuable waste correctly. From this point of view, they designed a holistic waste management system that tries to tackle these problems using modern Information Technology. The system consists of 2 parts; the first one is the software design and implementation including web-based waste management application and mobile application-based rewarding system; whereas the second one is the hardware design and implementation based on the Internet of Things (IoT) and Machine Learning technology.

This part can recognize and separate 3 kinds of waste automatically, namely glass bottles, plastic bottles, and metal cans. In normal operation, a user will scan a QR Code in front of the smart reward recycle bin using his/her smartphone. After scanning, he/she will be registered to a server and ready for collecting the points. The next step is to insert the valuable waste into the opening, one by one. The system will then capture the waste images and recognize them automatically. Based on valuable waste types, the point will be calculated and recorded in the user's account. The user can view the points on his/her smartphone. For waste management purposes, such as bin fullness monitoring or route calculation for garbage collection trucks, the web application comes into action. This smart rewarding recycle bin processing unit contains two main parts, namely the hardware and the software. The hardware part was designed to fit 3 standard 60-liter bins provided by the Saensuk municipal administration. These bins are prepared for 3 different types of valuable waste, which this smart rewarding recycle bin can separate automatically, namely plastic bottles, glass bottles, and metal cans. The system makes use of IoT technology to implement the hardware part of the system. The Raspberry Pi is used as a main processor controlling the operations of sensors and various electronic devices.

On the Raspberry Pi board, an Artificial Intelligence Model is installed and running. It will detect and separate the waste automatically, according to its properties. In the hardware setup, the Raspberry Pi is connected to the Internet via Wi-Fi which allows communication with a server and transmitting data via NETPIE service using MQTT protocol. The data payload includes the user data, the score data, and the amount of waste measured by ultrasonic sensors. Once the data has been received, it will be used in their developed mobile application and stored in a nonrelational database. The system allows the user to scan a QR Code and collect the points. The current condition for collecting the points is that the smartphone must connect to the Internet. The model for waste recognition must be improved with a large training dataset. The gap in their study is the solution for a stable and feasible Internet connection that must be tested and implemented. The frame of the bin will be redesigned and built with more durable and weatherproof materials. Saensuk City would benefit from the data collection and analysis.

*Poongodi et al.* [5] stated that people have become too dependent on technology owing to which the quantity of e-waste produced at the end of their life cycle has increased at a rapid pace. In later centuries, there would be an equivalent amount of carbon footprint by the smartphone sector as to that of the transportation industry. The approximate estimation of the total carbon footprint while taking the other existing abundant electronic gadgets is enormous. Considering the growing volume of e-waste, the possibilities of these non-biodegradable elements contaminating the atmosphere are towering. The e-waste generated in all these recent years is being taken seriously by various nations and effective steps are being taken to overcome this challenge.

The authors proposed an effective e-waste management technique by means of blockchain in the 5G scenario. The proposed solution tracks the e-waste produced and motivates people by providing incentives to them for channelizing the e-waste via agencies managed by the government that effectively dispose of the waste in an environment-friendly way. Henceforth, a partnership model is proposed for the implementation of this method which leads to an increase in jobs as well as proper organization of unplanned setup that is with a large amount of prospective potential. In this paper, they present a way to improve the situation of EWM in India. Their technique is based on smart contracts, developed using blockchain technology.

Bringing government agencies, consumers, and stakeholders on the same blockchain platform will lead to improved monitoring and higher transparency in the process. Blockchain will enable proper bookkeeping of the EEEs introduced in the market by different producers and retailers. This will enable smart contracts to clearly specify collection targets and penalize the appropriate party whenever required. They also proposed the inclusion of customers as members of this blockchain. Providing incentives to customers when they channelize their e-waste to the formal sector, can serve as the first step in reducing the dominance of the unorganized sector in EWM. They have also included collection centers as well as recycling units in their scheme. Smart contracts will help regulate the source and amount of e-waste collected, transported, and recycled throughout the process. Their proposed solution involves the whole of the users, dealers, and producers to finish the cycle from generating the goods to ensure that it is discarded correctly. Once the goods come to the e-waste center, the protected sum would be disbursed with an excess incentive to all the concerned stakeholders with those concerned goods. Still, even if after the ending of the life cycle of goods, it does not attain the e-waste center, the token fund would be surrendered.

Hence, this provides the contributing entities inspiration to ensure the goods arrive at e-waste center following its utilization. In this work, they have tried to present a robust, transparent, and safe solution to administer the e-waste problem using 5G Technology. By means of using the blockchain and smart contracts combined with 5G, they computerize the procedure of tracking a product until it arrives at the e-waste center and after that distribute the incentive amongst the entities taking part in the chain. The innovation of this solution is in the revelation of the big picture. By integrating a public-private partnership model connecting the government and private players, they could have an entire novel sector restoring the presently unorganized one encompassing unhygienic removal methods by ragpickers and kabbadiwalas (people who gather waste materials). At the present rate of electronic goods manufacturing, this can turn out to be a multi-billion-dollar business in a few years.

The incentive-driven model will give confidence to more and more entities ranging from producers to sellers and users to contribute to this system. As a result, a lot of jobs would be created. They could also separate the e-waste additionally to choose the objects which could be reused, and we could have a resale marketplace for such reused products. They have concluded that the environment would be saved from infectivity by e-wastes, the public would earn incentives from this scheme, a resale market could be started, and many jobs would be rolled out creating a win-win situation for each of the stakeholder entailed.

*Jadli and Hain* [2] stated that smart cities essentially combine the use of Information and communications technology (ICT) to provide services for better living conditions inside. The recent developments in the Internet of Things (IoT) opened vast opportunities for researchers and developers to implement various systems and applications in the field of Smart Cities and Intelligent Transportation Systems (ITS). It is a diverse topic of discussion with several application areas such as smart traffic management, smart streetlights or gas and water leak detection. Among these applications, Efficient Waste Collection (WC) is considered a fundamental service for helping maintain a healthier environment for the citizens while reducing operating costs. Their paper proposes a new architecture for a smart waste management system based on artificial intelligence techniques and focuses on the combined use of the Internet of Things (IoT) and surveillance systems as an assistive technology for high Quality of Service (QoS) in waste collection. By integrating deep learning techniques in this sector, important cost reductions can be made while maintaining optimal performance. Pattern recognition is a cognitive process that happens in our brain when we match some information that we encounter with data stored in our memory. In the context of machine learning, it is a technology that matches the information stored in the database with the incoming data. In the field of pattern recognition research, the method of using deep neural networks based on improved computing hardware recently attracted attention because of their superior accuracy compared to conventional methods.

Deep learning has been widely used and has become one Convolutional neural network (CNN) is a deep learning model that can achieve superb results in computer vision, natural language processing (NLP), and speech recognition because of its effective and efficient feature extraction capability on highly challenging datasets. In machine learning problems, the lack of labelled data can make supervised learning algorithms fail to build accurate classification models. Transfer learning has been developed to deal with such a lack of label problem. It aims to improve the performance of learning by transferring knowledge from several source domains to a target domain. Classifier-based knowledge transfer is a significant part of existing visual transfer learning techniques, and it has attracted much attention in recent years.

However, unlike the feature representation level knowledge transfer techniques, where only the training samples themselves in the source domain are adapted to the target learning framework, classifier-based knowledge transfer methods share the common trait that the learned source domain models are utilized as prior. knowledge in addition to the training samples when learning the target model. Instead of minimizing the cross-domain dissimilarity by updating instances' representations, classifier-based knowledge transfer methods aim to learn a new model that minimizes the generalization error in the target domain via provided training instances from both domains and the learned model. The use of technological advances in IoT in trash bins has been discussed and researched, but this solution was not adopted in Morocco due to very high costs and limited efficiency. They proposed a system based on artificial intelligence to detect, using solar-powered surveillance cameras, trash bins' current state by using image classification and object recognition. They added a new approach to a smart waste management system based on Deep learning and pattern recognition. This system can be integrated into existing solutions to help reduce costs and automate processes. as a perspective, we intend to collect real-world data to conduct the experiment and assess the generalization skill of the constructed model.

## 3 ARGUMENT AND DISCUSSION

This research study focuses on waste management in the city of Johannesburg. The literature review is formed from more than 8 peer reviewed articles. From the peer reviewed articles, all authors have adopted the waste management factor to help resolve the high levels of waste produced in their own countries and areas. Waste is a problem not only in South Africa but to the world in large. As countries are developing as well as the increasing number of people in this world, more waste is produced every single day.

From the literature review above most of peer reviewed articles demonstrated the use of systems that consists of sensors, CCTV cameras, and the GIS systems. The use of sensors in this case is a very vital factor because by using the sensors in the system it enables real-time collection of data. The sensors help in sensing and collecting the level data of waste inside the waste bins. This information is then communicated to the waste municipality centre in the form of an application. Waste trucks are sent to the area to empty the bins that need most of the attention. But the questions may arise for example: How reliable can the sensors be? What if they communicate the wrong information? In this paper, we make use of a redundant set of sensors in a way that both sensors to collect the information and if they both collect the same information it means the data is 99% percent correct. But when one sensor says the opposite then the trucks can go to two places instead of one place to make sure they pay attention to the al the waste bins in the area.

The are some studies of about 20% that focused on collecting waste bins house by the house as per a request. This paper focuses on a collective collection of waste because of the high population in the city of Johannesburg. Having waste collection from the house by the house will be a waste of time, costs, and resources. This paper focuses on reducing costs, time, and resources.

In some studies, about 40% of the literature review is approximately about CCTV cameras. Having CCTV cameras in the city of Johannesburg have its own disadvantages such as costs and safety. Including the CCTV cameras in the system will increase the costs of the system because CCTV cameras are expensive. Secondly having CCTV cameras in the streets of Johannesburg is risky because of the increasing level of crimes in the area. In the proposed solution cheaper cameras are used and placed at the top of the poles so that they cannot be accessed very easily. The cameras will make people aware that they are being watched in the case of crime and other people who might want to damage the embedded units. And again, cameras will help the waste municipalities to know the specific streets that have waste on the streets or in the public areas. Lastly, this will serve as a monitoring tool for the waste workers if they are working as expected.

This research paper proposes a waste management system that consists of Wireless Sensors, GPS, Monitoring centre, User application, and Education awareness.

## 4 PROPOSED SOLUTION

As the population grows, the waste produced by both humans and animals increases as well. The government has tried bringing in the waste municipalities to assist with waste. Waste Municipalities face various challenges as mentioned above. This paper is aimed at applying the ICT system in waste municipalities to manage waste smartly/effectively/efficiently. The below outlines solutions to helping waste municipalities manage waste. The smart waste management system will be able to manage waste in the most effective way which is superior to the normal traditional method. The proposed solution will consist of:

1. Smart waste bins – the smart bins embody a smart waste management model that contains sensors, microcontroller, GPS, and GSM. This model is placed under the lid of the smart waste bin. Considering the longevity, steel has been chosen over plastic. The steel hardware of the bin is chosen also because it will not be easy to pick and steal a heavy steel bin.
2. Wireless Sensors – Ultrasonic sensor that senses the level of waste in waste bins is used in this study. They are used to sense the level of waste in waste bins. The sensing, actuation, and control to describe and analyze a situation – in this case, the level of waste and make decisions based on the available data in a predictive or adaptive manner, thereby performing smart action. This sensor will be an indication if the smart waste bins are full or not. If they require attention or not. The sensors then send messages that the bin is full to the monitoring municipal center thereby saving costs and time of trips made by the waste collection trucks. Because of the various faults such as electrical faults, this study introduces a redundant system that consists of two same sensors that work the same way. When one sensor fails the other one will proceed where the other one left off. Sensors are the best technology as they gather real-time data, and they are reliable and consistent. Ultrasonic ranging module HC - SR04. This module can detect distances in the range of 2cm - 400cm. This module also includes ultrasonic transmitters, receiver, and control circuit.
3. Microcontroller - the ATmega 328P, which was used in this work by considering key specifications that meet the requirement for the hardware system. ATmega 328P is a low-power AVR 8-bit microcontroller that is based on advanced RISC architecture.
4. GPS – tracking system.
5. Camera – Each waste bin will be embedded with a camera (at the top of the pole) with motion sensors to make people aware that they are being watched in case if they would want to steal or damage the embedded units.
6. Power - The main power supply of the smart bin is a 9V DC battery.
7. GSM - internet connectivity to the cloud-based. Whenever the bin is full ultrasonic sensor will detect the waste level and will send the data to the zone control unit by GSM.
8. Monitoring center - The zone control unit monitors all the smart bins of its region and provides services as the bin requires. The zone control unit will receive all the data which will be sent by smart bin.
9. User application - The system provides a monitoring platform for the waste management institution to handle the alert records by creating orders for the garbage collectors/drivers which can be accessed via a mobile application system. The ultrasonic sensor that had been connected will measure the distance or the level of the waste and GSM will send those statuses to the control unit.

The proposed model contains the sensor, the microcontroller, GPS, and GSM will detect if the waste bin is full. If the level of waste is high the sensor model will communicate with the GSM to send the data via the internet to the application being used in the monitoring center. The data will be containing the GPS coordinates of the waste bin and the level of waste. The municipal waste monitoring center will then send out waste collectors to collect the waste using google maps to help find quicker routes.

## 5   EDUCATION AWARENESS

There will be centralized huge bins that will be placed in each region. These bins will consist of wireless sensors to sense the level of waste put inside and report back to the monitoring centre. The proprietary dumpster can "talk" to the office of waste collection when it is filled up thereby saving cost and time of trips made by waste collectors. These sensors can also help the company forecast dumpster filling patterns [8]. Each bin will consist of a GPS tracker with a unique parameter. When one specific bin is full and the sensor can attest to it both the sensor results and the GPS coordinates will be sent to the monitoring centre where waste municipalities will go directly to that bin to collect all the waste for further processing. The next aspect is the transparent user application that will be application that is free to download and to use.  The advantage of this application is that not 98% of the communities have smartphones which will enable easy use and access to this application. This application will be user-friendly and can be used on all operating systems. This app enables members of the community in knowing which bins are full, which bins are empty, and the location of each bin. This will help in reducing litter as there will not be any excuse for having full bins that are not attended to.

   The last aspect is the educational awareness that needs to spread to all corners of the world. Traditional ways of educating people about waste management are ancient, this paper will enable user applications to have education awareness which will be easy to read and access. This application will not only tell the communities about waste but will also teach them about waste management. The disadvantages of waste, littering, and the health problems that come with waste. This application will enable people to learn how to report, deal, and lastly make their region waste free. This application may involve messages such as keeping garbage bins in good locations; buying goods without unnecessary packaging; returning packaging materials to production sites; returning products for collective disposal; and recycling items such as used furniture, electrical, and electronic appliances. The commencement of a community environmental education awareness program is essential to rapidly educate the public and facilitate the development of environmentally friendly community waste behaviour. Conclusion This model is highly effective, and it will produce a cleaner, hygienic environment.

## 6   Conclusion

The problem of waste has been proven to be a global problem as per the peer reviewed articles from multiple countries. This paper proposed a system that solves the waste problem we are facing in the city of Johannesburg entirely. The proposed system detects and reports full-capacity bins so that they can be attended to with immediate effect, leaving the bins available for public use. The proposed system also helps the waste Municipality reduce time and costs to produce smart working habits. The proposed solution has proven to work efficiently, redundant, and reliable. Overflowing waste bins is a thing of the past. The user application has proven that awareness can be spread over the city with only a click on the application. The proposed solution can be adopted in other areas and cities as waste is a global problem.

### Acknowledgments

The authors would like to thank the Tshwane University of Technology for financial support. The authors declare that there is no conflict of interest regarding the publication of this paper.